\documentstyle[11pt,newpasp,twoside,epsf]{article}
\def\edcomment#1{\iffalse\marginpar{\raggedright\sl#1\/}\else\relax\fi} 
\reversemarginpar

\begin{document}
\title{The mass distribution in early type disk galaxies}
\author{Edo Noordermeer, Thijs van der Hulst}
\affil{Kapteyn Institute, PO Box 800, 9700 AV Groningen, The
Netherlands}
\author{Renzo Sancisi}
\affil{Osservatorio Astronomico, Via Ranzani 1, 40127 Bologna, Italy 
\& Kapteyn Institute, PO Box 800, 9700 AV Groningen, The Netherlands} 
\author{Rob Swaters}
\affil{Johns Hopkins University, 3400 N. Charles St., Baltimore MD
21218, U.S.A. \& Space Telescope Science Institute, 3700 San Martin
Dr., Baltimore, MD 21218, U.S.A.} 

\begin{abstract}
We are studying the mass distribution in a sample of 50 early type
spiral galaxies, with morphological type betweens S0 and Sab and
absolute magnitudes M$_B$ between -18 and -22; they form the massive
and high-surface brightness extreme of the disk galaxy population. Our 
study is designed to investigate the relation between dark and
luminous matter in these systems, of which very little yet is known.

From a combination of WSRT H\/{\sc i} observations and long-slit
optical spectra, we have obtained high-quality rotation curves. The
rotation velocities always rise very fast in the center; in the 
outer regions, they are often declining, with the outermost measured
velocity 10-25\% lower than the maximum.  

We decompose the rotation curves into contributions from the luminous
(stellar and gaseous) and dark matter. The stellar disks and bulges
always dominate the rotation curves within the inner few disk scale
lengths, and are responsible for the decline in the outer
parts. As an example, we present here the decompositions for
UGC~9133. We are able to put tight upper and lower limits on the
stellar mass-to-light ratios. 
\end{abstract}

\section{Introduction}
Rotation curves, in particular from H\/{\sc i} data extending well
outside the optical disks, have been proven in the last 30 years to be
a useful tool for the study of dark matter in disk galaxies. A recent
review on the observations and properties of rotation curves has been
given by Sofue \& Rubin (2001). H\/{\sc i} observations of early type,
massive disk galaxies are however rare. Broeils (1992) showed a
compilation of all high-quality H\/{\sc i} rotation curves known to
that date; only 4 out of 23 galaxies had maximum rotation velocity
$V_{\mathrm max} > 250$ km/s. Since then, many more low- and
intermediate mass galaxies have been studied in H\/{\sc i} (De Blok,
McGaugh \& Van der Hulst 1996, Swaters 1999, Verheijen \& Sancisi
2001, Cot\'{e}, Carignan \& Freeman 2000), but little progress has
been made on the high-mass side. The Universal Rotation Curve of
Persic, Salucci \& Stel (1996) was based on 1100 rotation curves, but
only 2 were of type Sab or earlier.  

In this study, we address specifically the relation between luminous
and dark matter in massive, early-type spiral galaxies. We investigate
whether these systems follow the general trends that exist in later
type galaxies (e.g. Tully-Fisher, dark matter content
vs. morphological type or surface brightness, etc.), or that they form 
a distinct class of galaxies with different characteristics. 

\section{Sample selection and Observations}
The H\/{\sc i} observations are taken from the WHISP survey. WHISP
galaxies were selected from the UGC on 
the basis of position on the sky ($\delta > 20\deg$), angular size 
($D_{25} > 1.5\arcmin$) and H\/{\sc i} line flux ($F_{\mathrm H \/
{\sc i}} > 20 {\mathrm mJy}$). These criteria ensure that the galaxies
are well resolved, and that the signal-to-noise ratio of the
observations is sufficient. 

The total survey contains about 80 galaxies with morphological type
between S0$^-$ and Sab. From this subsample, we choose by eye the
approximately 50 galaxies that have a well-defined velocity field
showing ordered rotation. 

We have obtained additional long-slit H$\alpha$ spectra at high angular
resolution from the INT on La Palma, to circumvent the problem of
beam-smearing in the H\/{\sc i} data. The stellar mass distribution is 
determined from multi-color broad-band optical images taken with the
JKT on La Palma.  

\section{Shapes of the rotation curves}
In fig.~1, we show rotation curves of 9 galaxies, derived using a
tilted-ring fit to the H\/{\sc i} velocity fields. 
\begin{figure}[ht]
 \plotone{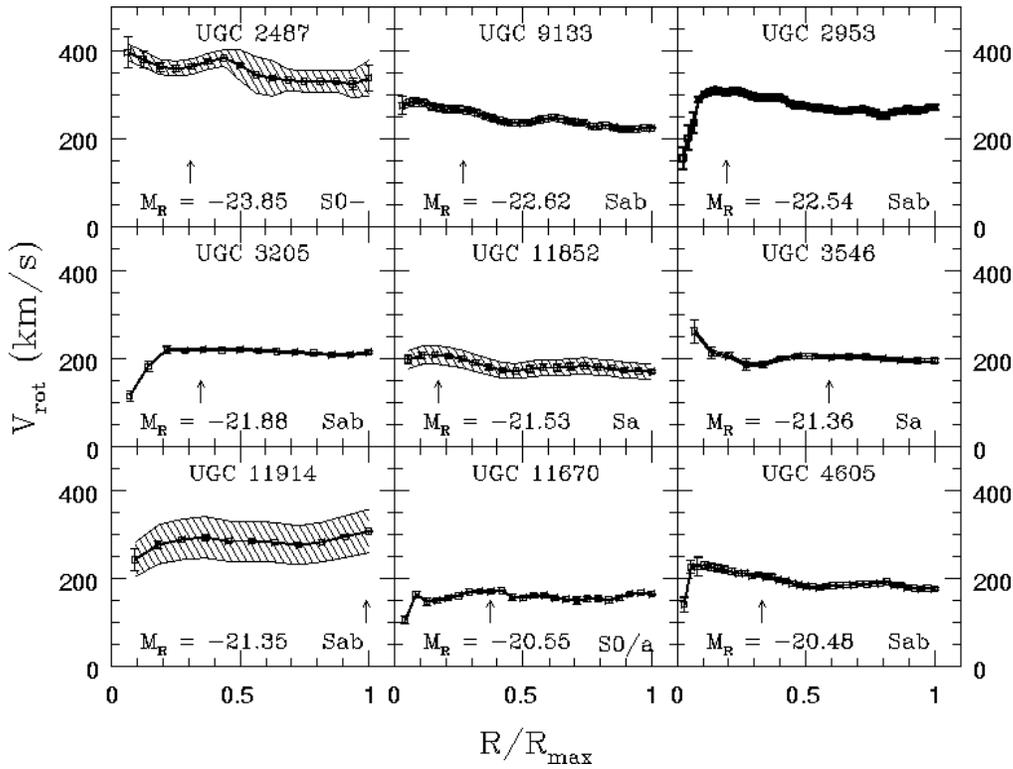}
 \caption{\small{H\/{\sc i} rotation curves of 9 galaxies from the
 sample. All radii are scaled to the radius of the last measured
 point. The mean, inclination-corrected rotation velocities are
 indicated by the symbols, connected by the thick line. The errorbars
 show a combination of statistical errors and contributions from
 asymmetries in the velocity field. The hatched regions indicate the
 errors induced by uncertainties in the inclinations; they are
 often too small to be visible here. The arrows indicate the extent of
 the stellar disk (D$_{25}/2$).}} 
\end{figure}
No H$\alpha$ observations have been used for these
curves. This figure shows some properties that are typical for the
rotation curves of this type of galaxies:
\begin{itemize}
 \item The rotation curves rise extremely fast in the inner
 regions. Often, the maximum velocity is reached within one H\/{\sc
 i}-beam. H$\alpha$ observations are clearly needed to resolve the
 central velocity gradient.  
 \item In the outer regions most rotation curves show a slow but 
 systematic decline. In some cases the outer velocity $V_{\mathrm 
 out}$ is less than 80\% of the maximum velocity. This decline can 
 {\em not} be explained as the result of warping in the outer
 regions. The uncertainties in the inclination and position angle are
 usually well understood; the resulting errors in the velocities are
 indicated by the hatched regions. The fact that so many rotation
 curves are declining, also argues against warps; if warps have no
 preferential direction, they should equally frequently produce rising
 rotation curves.

 Casertano \& Van Gorkom (1991) argued that declining rotation curves
 mostly occur in galaxies with a compact light distribution,
 i.e. small scale lengths. However, some of our galaxies with
 declining rotation curves are not compact at all. UGC~9133 for
 example, having one of the most strongly declining rotation curves in 
 the sample, has a disk scale length of 9.3 kpc. Our results seem to
 indicate that the majority of early-type disk galaxies have rotation 
 curves that are more or less declining in the outer parts. 
 \item The maximum rotation velocities are large: $200 < V_{\mathrm max} <
 400\, {\mathrm km/s}$. This may in principle be due to selection
 effects in our sample definition. We will investigate this issue
 further in the future. 
\end{itemize}

The rotation curves presented in fig.~1 look remarkably different from
the ones of later-type galaxies, which usually rise slowly and then
become flat at large radii. Persic et al. (1996) found, from a large
sample of later-type galaxies, strong correlations between rotation
curve shape and amplitude on one hand and other structural parameters
such as luminosity, surface brightness, size, etc. on the other
hand. For a good understanding of the relation between dark and 
luminous matter, it is important to know whether these relations can
simply be extrapolated to our bright early-type galaxies, or that they
need further fine-tuning. We will investigate this at a later stage. 

\section{Mass modeling: UGC~9133}
The major goal of our study is to get a detailed picture of the mass
distribution in individual galaxies. We decompose the rotation curves
into contributions from stellar bulges and disks, neutral gas and dark
matter. To illustrate the method, we show here the analysis for
UGC~9133. Some basic parameters of this galaxy are given in table~1,
and an R-band image from the JKT is shown in fig.~2. The H\/{\sc i}
observations are shown in fig.~3.
\begin{figure}[ht]
 \begin{minipage}[c]{5cm}
  \centering
  \small{
   \begin{tabular}{ll} 
    \multicolumn{2}{c}{Table 1. UGC~9133: basic data} \\ \hline 
    NGC			& 5533 \\
    RA (J2000)		& 14$^h$ 16$^m$ 7.76$^s$ \\
    Dec (J2000)		& 35\deg 20' 38.3'' \\
    Type  		& Sab \\
    Distance		& 54 Mpc \\
    M$_B$		& -21.22 \\
    H\/{\sc i} mass	& $2.7 \cdot 10^{10} M_{\sun}$ \\
    Disk scale length 	& 9.3 kpc \\
    Bulge-Disk ratio	& 0.74 \\ \hline
    \multicolumn{2}{l}{\footnotesize{Numbers are based on a Hubble}}
  \\  
    \multicolumn{2}{l}{\footnotesize{constant of H$_0$ = 75
  km/s/Mpc.}}\\ 
    \multicolumn{2}{l}{\footnotesize{The scale length and B/D-ratio are}}
  \\
    \multicolumn{2}{l}{\footnotesize{measured on the R-band image
  shown}} \\
    \multicolumn{2}{l}{\footnotesize{in fig.~2.}}
   \end{tabular}
        }
 \end{minipage}
 \hspace{0.6cm}
 \begin{minipage}[c]{7.5cm}
  \centering
  \epsfbox{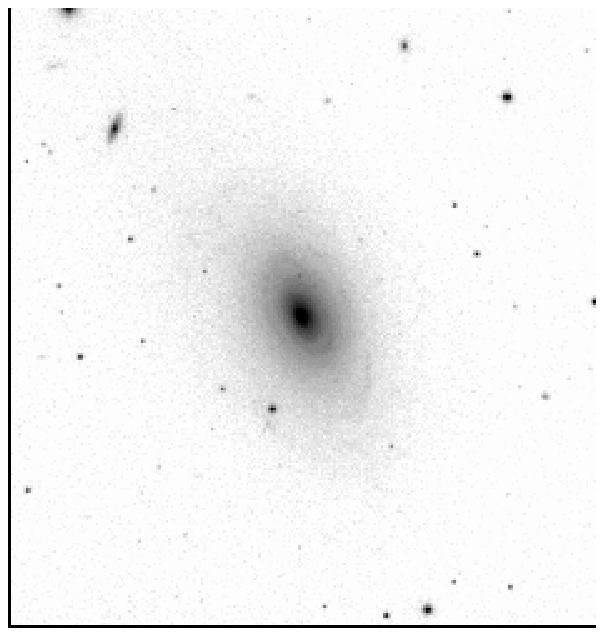}
  \caption{\small{UGC~9133: R-band image}} 
 \end{minipage}
\end{figure}

To account for the contribution from the stars to the rotation curves,
we extract a surface brightness profile from the optical image, by
azimuthally averaging the intensities over concentric elliptic annuli.
A bulge-disk decomposition is performed by fitting the sum of a
Sersic-bulge and an exponential disk to the profile. Each component is
assigned a mass-to-light ratio (M/L) which is assumed to be constant
with radius. We then follow the procedures given by Casertano (1983)
to calculate the contribution from both components to the rotation 
curve. A similar procedure is followed with the H\/{\sc i} density 
distribution. To account for the presence of primordial Helium, the
H\/{\sc i} surface densities are multiplied by 1.43. 
\begin{figure}[ht]
 \plotone{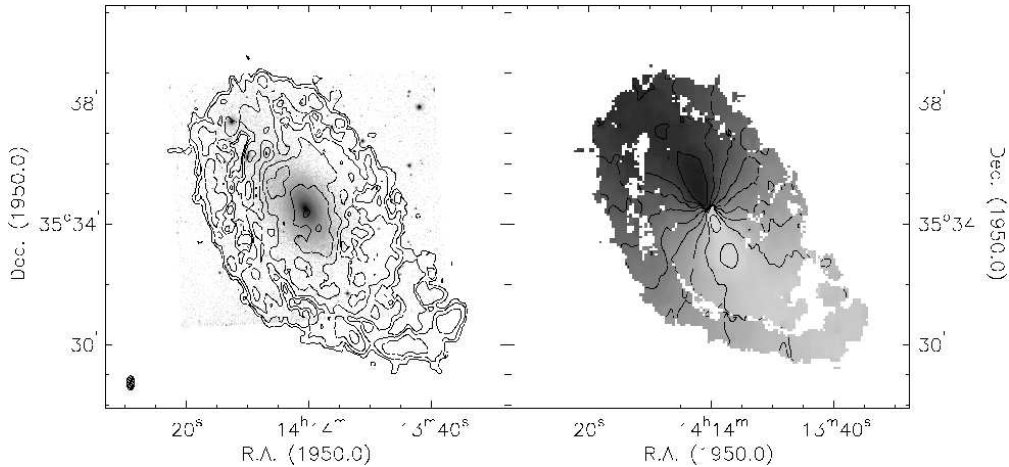}
 \caption{\small{{\bf left:} H\/{\sc i} contours overplotted over the R-band
 image from fig.~2. Thin contours are 1,2,4,...$\cdot \sigma$, with
 $1 \sigma = 2.5 \cdot 10^{19}$ atoms/cm$^2$. The FWHM beamsize of the
 H\/{\sc i} observations is indicated as the ellips in the bottom left
 corner. {\bf right:} H\/{\sc i} velocity field. Contours are spaced
 50 km/s, from 3650 to 4000 km/s. Dark shaded area indicates the
 receding side.}} 
\end{figure}

Although the rotation velocities in the inner parts can usually be 
explained by the luminous matter only, a dark matter halo is needed to
explain the rotation curve in the outer regions. This is modelled as a
pseudo-isothermal sphere with 2 free parameters: the central density
$\rho_0$ and the core radius $R_c$. The full model thus consists of 4
free parameters: 2 M/L-ratios and 2 halo parameters. 
\begin{figure}[htb]
 \plotone{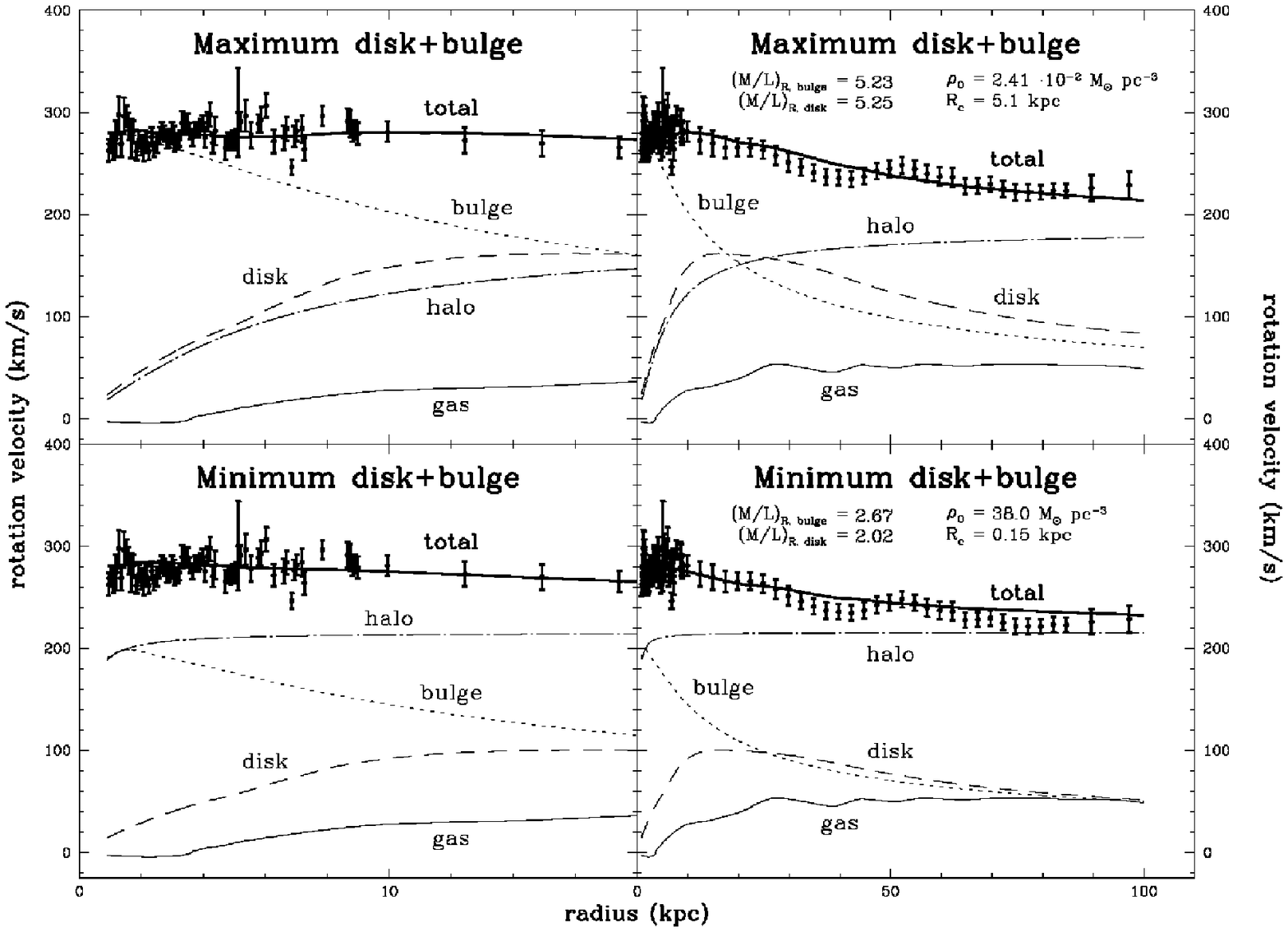}
 \caption{\small{Mass models for UGC~9133. Right hand panels show the full
 rotation curves, left hand panels are a blow-up of the inner
 parts. The upper panels show the models with maximal contribution of
 bulge and disk to the rotation curves, lower panels show the minimal
 contributions. Symbols with errorbars denote the measured rotation curves,
 from H$\alpha$ ($r < 10 {\mathrm kpc}$) and H\/{\sc i} ($r > 10
 {\mathrm kpc}$). The separate curves show the contributions of bulge
 (dotted), disk (short dashed), gas (thin solid) and dark halo
 (dot-dashed lines). The M/L-ratios for bulge and disk and the halo
 parameters are given in the right-hand panels. The thick solid line
 shows the sum of the separate components.}}
\end{figure}

As in later-type galaxies, there is a degeneracy between the stellar
M/L-ratios and the halo parameters. The maximum M/L-ratios are reached
in models where the bulge and disk account fully for the observed
rotation velocities in the inner regions, with the dark halo only
contributing at larger radii ('maximum bulge+disk', upper panels in 
fig.~4). In this case however, the steep rise of the rotation curve in
the central regions and the decline at larger radii also allow us to
put tight lower limits to the M/L-ratios for bulge and disk ('minimum
bulge+disk', lower panels in fig.~4), thereby fixing the stellar
M/L-ratios to within a factor of 2. This is in contrast with
late-type and LSB galaxies, where usually no lower limits can be 
given (Swaters 1999, De Blok \& Bosma 2002). 

Only by invoking exotic, highly concentrated dark matter models, can
we decrease the stellar M/L-ratios further. Already our minimum
bulge+disk model shown in fig.~4 has an extremely concentrated dark
halo, at odds with current theories of structure formation. Alam,
Bullock \& Weinberg (2002) use numerical simulations to predict the
densities and concentrations of dark matter halos. They 
use the quantity $\Delta_{V/2}$, and derive values between $10^5$ and
$10^6$ for the range of galaxies in our sample. Indeed, our maximum
bulge+disk model has \mbox{$\Delta_{V/2} = 1.1 \cdot 10^5$}; however,
our minimum bulge+disk model has \mbox{$\Delta_{V/2} = 1.2 \cdot
10^8$}, orders of magnitudes too high. We conclude that the M/L-ratios
for the stars must be close to the maximum values shown in the upper
panels of fig.~4 for any realistic dark halo.   

\section{Conclusions}
We study the rotation curves and mass distribution in a sample of 50
early-type disk galaxies (S0 to Sab); this part of the galaxy
population has not well been studied yet. 

The rotation curves rise extremely fast in the center, and are often
declining at large radii. Their shapes can only be explained if the
stellar bulge and disk dominate the gravitational potential in the
central regions. Dark matter is needed to explain the observed
rotation velocities in the outer regions. A range of stellar
mass-to-light ratios can be used to fit the data, but we are able to
constrain the M/L-values with tight upper and lower limits.


\begin{references}
\reference{Alam, S.M.K., Bullock, J.S., \& Weinberg, D.H. 2002, \apj,
572, 34}
\reference{De Blok, W.J.G., McGaugh, S.S., \& Van der Hulst,
J.M. 1996, \mnras, 283, 18}
\reference{De Blok, W.J.G., \& Bosma, A. 2002, \aap, 385, 816} 
\reference{Broeils, A.H. 1992, PhD thesis, Rijksuniversiteit
Groningen}
\reference{Casertano, S. 1983, \mnras, 203, 735}
\reference{Casertano, S., \& Van Gorkom, J.H. 1991, \aj, 101, 1231}
\reference{Cot\'{e}, S., Carignan, C., \& Freeman, K.C. 2000, \aj,
120, 3027}
\reference{Persic, M., Salucci, P., \& Stel, F. 1996, \mnras, 281, 27} 
\reference{Sofue, Y., \& Rubin, V. 2001, \araa, 39, 137}
\reference{Swaters, R.A. 1999, PhD thesis, Rijksuniversiteit
Groningen} 
\reference{Verheijen, M.A.W., \& Sancisi, R. 2001, \aap, 370, 765}


\end{references}
\end{document}